# Safeguarding Patient Trust in the Age of AI: Tackling Health Misinformation with Explainable AI

Authors:
Ms. Sueun Hong[a], Ms. Shuojie Fu[b,c], Dr. Ovidiu Serban[b], Ms. Brianna Bao[b], Dr. James Kinross [b,c], Prof. Francesa Toni [b,c], Dr. Guy Martin [b], Dr. Uddhav Vaghela[b]

[a] NYU Langone Health, Department of Population Health, 180 Madison Ave, New York, United States
[b] The Evidence Company Ltd., United Kingdom
[c] Imperial College London, United Kingdom

## Abstract

AI-generated health misinformation poses unprecedented threats to patient safety and healthcare system trust globally. This white paper presents an explainable AI framework developed through the EPSRC INDICATE project to combat medical misinformation while enhancing evidence-based healthcare delivery. Our systematic review of 17 studies reveals the urgent need for transparent AI systems in healthcare. The proposed solution demonstrates 95% recall in clinical evidence retrieval and integrates novel trustworthiness classifiers achieving 76% F1 score in detecting biomedical misinformation. Results show that explainable AI can transform traditional 6-month expert review processes into real-time, automated evidence synthesis while maintaining clinical rigor. This approach offers a critical intervention to preserve healthcare integrity in the AI era.

---

## 1. The Misinformation/Infodemic Crisis in Healthcare and Implications for Patient Safety & Health System Trust

With the advent of generative artificial intelligence, a new era of sophisticated but potentially dangerous medical disinformation has begun, endangering patient safety and international health systems. AI generative platforms can potentially improve healthcare delivery, but their development and application must be made carefully to prevent the provision of medically hazardous content that could undermine confidence in evidence-based medicine.

The World Health Organisation (WHO) has recognised health misinformation as a global threat, coining the term "infodemic" to describe the harm the proliferation of health misinformation and disinformation may cause(Organization, 2020). According to WHO's definition, an infodemic represents " false or misleading information in digital and physical environments during a disease outbreak" that "causes confusion and risk-taking behaviours that can harm health" and "leads to mistrust in health authorities and undermines the public health response."(Organization, 2024) This phenomenon has evolved from a minor concern during health emergencies to a significant international threat.

Unlike traditional forms of health misinformation, AI-generated content can produce deceptively detailed information that appears credible even to healthcare professionals. The British Journal of Psychiatry warns that "misinformation created by generative AI about mental illness may include factual errors, nonsense, fabricated sources and dangerous advice," highlighting how AI systems can generate content that mimics legitimate medical guidance while containing potentially fatal recommendations. (Monteith et al., 2024)

The worldwide scale of this threat is demonstrated by several alarming examples from many healthcare areas, including vaccine misinformation, fraudulent cancer cures, and risky medical procedures. The U.S. Department of Health and Human Services notes that "the growing number of places people go to for information has made it easier for misinformation to spread at a never-before-seen speed and scale," with false information consistently outpacing verified medical advice across digital platforms. (Services, 2024)This acceleration is especially risky in healthcare settings where inaccurate or delayed information can directly impact patient outcomes and clinical decision-making.

The population to which misinformation is spreading is additionally essential to note. The crisis is most acute in healthcare systems characterised by low public trust, fragmented care delivery, or limited access to authoritative medical sources. Patients increasingly turn to online sources for health information in these environments, making them particularly vulnerable to easily accessible AI-generated misinformation. Research demonstrates that medical misinformation spreads faster than verified information due to algorithmic amplification on social media platforms, prioritising engagement over accuracy, and the psychological tendency for sensational or fear-inducing content to achieve broader reach than measured scientific communication. (Carletto et al., 2025) The implications for patient safety extend beyond individual harm to systemic threats to public health infrastructure. Unlike traditional public health threats that typically have geographic or demographic boundaries, the infodemic represents a truly global health risk that transcends traditional epidemiological patterns and requires coordinated international response mechanisms.

This white paper addresses the urgent need for explainable health AI systems as a critical intervention to safeguard patient trust and combat health misinformation in an era where artificial intelligence serves as the primary vector for sophisticated medical misinformation and the most promising technological solution for its detection and mitigation. The stakes could not be higher: the integrity of global health systems, the safety of patients worldwide, and the fundamental relationship between healthcare providers and the communities they serve all depend on our collective ability to harness AI's potential while neutralising its capacity for harm.

## 2. The Shortcomings of Current AI Systems in Healthcare and our Approach (EPSRC INDICATE Project)

Various AI-driven healthcare providers aim to leverage artificial intelligence to enhance service delivery. However, several challenges remain unsolved or even worsened in some scenarios.

In the current healthcare technology landscape, numerous platforms (e.g. Hippocratic AI) (*Hippocratic AI*; *Interactly AI*)offer multi-agent systems to assist with medical task execution. These architectures typically consist of several medical agents managed by a primary agent operating under a waterfall-style workflow. Such a structure is inherently vulnerable to cascading errors, as inaccuracies introduced in early stages may propagate through the system and lead to compounded failures. This issue is particularly found in platforms (*Hyro AI*) that rely on keyword-based entity recognition to guide decision flow (e.g. Hyro AI), which can result in misleading outcomes. Human-in-the-loop recalibrations are often included to mitigate these errors; however, this reliance on manual oversight undermines the intended gains in efficiency. Another critical limitation of multi-model architectures lies in maintaining consistency across individual models. It is challenging to ensure all constituent models are updated uniformly and coherently with each other, increasing the risk of contradictory outputs.

Furthermore, some platforms (*Flagler Health*; *Quadrant Health*; *WayFinding*) integrate non-clinical functionalities within the same system (e.g. Flagler Health, Quadrant Health, WayFinding), such as providing parking information and clinician calendar checking. These logistic features must be compartmentalised appropriately to avoid interfering with the precision and reliability required for medical-related tasks. Additionally, various platforms (*Care Navigation*; *Notable Health*; *WayFinding*) act as medical query receptionists (e.g. Care Navigation, Notable Health), providing clinical guidance tailored to individual organisations. Many platforms (*Evidium*; *Qualified Health AI*) that claim to provide evidence-based guidance do not disclose the methodologies used or the frequency of evidence updates (e.g. Evidium; Qualified Health AI). As an example of an evidence-supported AI healthcare system, we proposed a semi-automated framework for medical guideline creation (the INDICATE project).

In the UK, medical guidelines are evidence-based recommendations that guide healthcare professionals in caring for individuals with specific conditions. The National Institute for Health and Care Excellence (NICE) is the leading organisation responsible for creating and issuing these guidelines in England and Wales, offering advice on prevention, treatment across primary and secondary care, and specialised healthcare services.

In a conventional setting, medical guideline creation involves several sequential steps followed by many organisations producing systematic reviews: (1) formulating key questions based on established guidelines; (2) systematically identifying all relevant medical publications within a predetermined time frame; (3) assembling subject-matter experts to review the retrieved literature and select the most pertinent studies; (4) determining which publications meet the criteria for inclusion as medical guideline references; and (5) synthesising findings by having experts compose summaries of selected studies that directly address the predefined questions. This is a time-consuming and labour-intensive process that often involves efforts from dozens of healthcare professionals. Our proposed methodology automates some of these steps by using natural language processing (NLP) techniques and large language models (LLMs) for information retrieval, quality checking and evidence summarisation. Meanwhile, human-in-the-loop checks remain to ensure the accuracy and quality of the outcome.

## 3. Methodology of the White Paper - Systematic Review

A systematic review explored the current state of AI-driven health misinformation. Our systematic review follows the Preferred Reporting Items for Systematic Reviews and Meta-Analyses (PRISMA) guidelines(Moher et al., 2009). We conducted comprehensive searches across multiple databases, including MEDLINE, Embase, PubMed and grey literature sources. We looked at 17 studies through these searches.

Search terms combined controlled vocabulary and natural language terms related to: (1) artificial intelligence AND health misinformation, (2) explainable AI AND healthcare trust, (3) AI transparency AND patient safety, and (4) trustworthy AI AND clinical decision-making. The inclusion criteria for the present review include the following:

- Peer-reviewed studies published between 2020 and 2024
- Research addressing AI-generated health misinformation detection, prevention, or mitigation
- Studies on explainable AI applications in healthcare contexts
- Government policy documents and regulatory frameworks for AI in healthcare
- Empirical studies measuring trust in AI health systems

We excluded the following**:**

- Studies not available in English
- Conference abstracts without full papers
- Studies focusing solely on technical AI development without healthcare applications
- Research predating the widespread adoption of generative AI

In our proposed framework for creating medical guidelines, we replaced some of the traditional steps with NLP techniques and LLMs to improve the efficiency and accuracy of the process.

First, a database was created by crawling publishers, including PubMed, CENTRAL and CDSR, for open-source, peer-reviewed medical publications related to breast cancer. This crawling process could be performed once every month or every fortnight to ensure that the latest publications are continually updated in the database. After several preprocessing steps, including cleaning and chunking the text, an encoder was used to extract numeric representations of the text chunks. These representations, called embeddings, were stored in a vectorised knowledge base. Several encoders (Lee, 2024; NeuML/pubmedbert-base-embeddings, 2025) were used in this step, and the results were evaluated using reference data.

When a question was received, we extracted the question embedding using an encoder that projects into the same hyperspace as the articles. The question embedding was then passed into the knowledge base for similarity search, and the top K most relevant article passages were retrieved using various similarity comparison metrics. After this, the retrieved passages were reranked by a Cross-Encoder (Lee, 2024; NeuML/pubmedbert-base-embeddings, 2025) to achieve a more precise ranking. The retrievers were fine-tuned on independent datasets, so the final selection is not biased in any way. The human experts then joined the process to read the retrieved passages and assign relevancy and curation scores to each publication. An automated veracity check pipeline also assigned a veracity score to each publication. The ones with low veracity scores would be less weighted or eliminated from the collection.

The selected collection of publications was used as input to an LLM, which was instructed to synthesise an answer to the initial medical question using the publications as evidence. The LLM was forbidden to use prior knowledge so that hallucination could be reduced to a great degree. The outputs were reviewed by a team of clinicians, who ranked them by clarity and usefulness.

## 4. Proposed Solution and Results from the INDICATE Project: Trustworthy AI for Navigating Care

We conducted a multi-stage evaluation of our AI-enabled clinical evidence synthesis framework, designed to automate and accelerate the creation of trustworthy, explainable health information. Using the NICE NG101 guideline ('Early and locally advanced breast cancer: diagnosis and management') on breast cancer as our reference, we benchmarked the system's ability to retrieve, validate, and synthesise evidence aligned with real-world medical questions.

**Stage 1: Semantic Evidence Retrieval Performance**

**Objective:** Evaluate the system's retrieval precision and recall when surfaced with complex medical queries.

- **Clinical Reference Benchmark:**

  We selected four breast cancer–related clinical questions derived from NICE Guideline NG101. Each question was associated with a curated set of relevant publications from NICE's internal review process, which served as a gold standard.

- **Retrieval Configuration:**

> We tested multiple combinations of domain-specific encoder models and similarity metrics to identify the optimal top-K retrieval performance.

- **Results:**
    - On the NICE-derived test set, our pipeline could classify both included and excluded studies, despite never having seen this literature before.
    - 95% recall was achieved on a synthetic benchmark collaboratively developed with NICE, demonstrating high coverage across structured clinical questions and diverse biomedical sources.
    - Error Analysis: Remaining omissions were due to either:
        - Semantic misclassification, such as failing to identify Pembrolizumab as a chemotherapy agent; or
        - Data incompleteness, where certain publications were not ingested due to paywall restrictions.

This RAG (Retrieval-Augmented Generation) pipeline dramatically reduces the time and resource burden of traditional evidence review, collapsing a 6-month, 20-person expert process into seconds—with real-time updating capabilities.

**Stage 2: Biomedical Answer Synthesis Quality**

**Objective:** Benchmark the output quality of LLM-generated clinical answers using blinded clinician review.

- **Methodology:**
    - A single clinical question was paired with three supporting biomedical publications.
    - Eight open-source LLMs (including variants of LLaMA, Mistral, and Claude) were prompted to generate evidence-grounded answers.
    - Responses were anonymised and embedded into a blinded questionnaire, distributed to a panel of clinicians instructed to select the top three answers based on relevance, clarity, and clinical rigour.

This enabled an unbiased evaluation of model performance in medical synthesis tasks, grounded in real-world clinical interpretation.

**Advanced Verification and Trustworthiness Models**

To further ensure the reliability and transparency of generated outputs, we integrated two novel AI components:

- **PubGuardLLM – Biomedical Trustworthiness Classifier (Chen et al., 2025)**

    Developed to detect misinformation and low-quality publications, PubGuardLLM achieves:

    - 76% F1 score across multiple biomedical datasets

- > 80% recall in detecting research fraud - without external supervision, indicating strong generalisation and standalone robustness.
- **argLLM – Structured Explainability & Claim Validation Model (Freedman et al., 2025)**

  Integrated to provide transparent reasoning and verifiable outputs, argLLM enables structured, auditable answer explanations with:

  - 90%+ F1 across multiple benchmark datasets
  - Support for multi-source claim validation and contradiction detection, critical for guideline-grade outputs.

Deploying such trustworthy AI solutions in UK healthcare requires careful navigation of an evolving regulatory landscape designed to optimise access, promote equity, and reduce system burden. Organisations developing explainable, personalised AI that empowers healthcare decision-making must align with multiple regulatory frameworks while maintaining their mission of transforming healthcare delivery through evidence-based insights. The UK has adopted a principles-based regulatory approach through collaborative frameworks that particularly benefit organisations prioritising trustworthiness and transparency in healthcare AI, with the AI and Digital Regulations Service coordinating between NICE, CQC, MHRA, and the Health Research Authority to create comprehensive guidance for AI adopters and innovators. (Service, 2024)

Organisations focusing on actionable, comprehensive healthcare solutions can leverage NICE's evaluation pathways to demonstrate clinical value and secure NHS adoption pathways. The NHS has established clear expectations for AI transparency and accountability, requiring all AI systems to be logged and regularly reviewed yearly, with transparency and public disclosure ensured. These ongoing compliance obligations favour explainable AI architectures and create natural alignment opportunities for organisations committed to transparency in their AI development processes.

Healthcare AI solutions that facilitate shared decision-making with trusted, evidence-based insights directly address the UK's regulatory emphasis on patient empowerment and healthcare accessibility. AI solutions must demonstrate how they enhance rather than burden clinical workflows, providing actionable insights that support evidence-based decision making without increasing administrative complexity for healthcare providers. Healthcare AI that bridges pharmaceutical innovation with patient care must navigate clinical and pharmaceutical regulatory requirements while focusing on patient outcomes and system efficiency in biopharma collaborations. Regulatory frameworks increasingly emphasise patient-centred outcomes, requiring AI solutions to demonstrate improved patient experience throughout the care continuum while maintaining privacy and safety standards.

The UK's evolving regulatory landscape creates clear pathways for deploying trustworthy AI solutions that optimise access, promote equity, and reduce the burden on the healthcare system. Success requires positioning trustworthiness, evidence-based decision making, and patient-centric care as core design principles rather than compliance afterthoughts. Organisations that embrace regulatory requirements as competitive advantages while focusing on explainable, personalised AI can leverage the UK's principles-based approach to create sustainable growth opportunities. By aligning with the vision of transforming the NHS into the world's most AI-enabled health system, healthcare AI solutions can bridge the gap between information and action while empowering trust in every health decision across the care continuum.

## 5. Impact and Benefits to Government

The UK government has already recognised that AI poses threats of misleading information and recognises the need to ensure AI-generated content is trusted and safe (UK Government Department for Science, 2024). The UK government must treat health misinformation as a public health threat requiring a coordinated national response. There has already been a case where a patient in London was falsely diagnosed with diabetes and heart disease. The NHS responded to the incident by limiting the use of supervised AI, which restricts innovation in the healthcare system.

Our review of existing literature and technologies suggests that rather than restricting the use of AI, the UK government should continue to embrace the possibilities of AI in healthcare. According to the British government's "10-Year Health Plan", the UK should continue using AI to reduce administrative burden and empower patients. This, however, will only be achieved by mandating transparent standards for AI health applications. Without government intervention, the health misinformation crisis will only continue to intensify and create cases similar to the falsely diagnosed diabetes patient. Successfully creating regulations for AI in healthcare will allow the UK to position itself as a leader in AI health regulations, a benchmark for international investment and strengthen the UK's position in global health diplomacy.

## 6. Ethical and Legal Considerations

Integrating AI into healthcare systems "raises significant ethical and legal challenges," requiring attention to key ethical principles—autonomy, beneficence, non-maleficence, and justice (Beauchamp & Childress, 2024). From a legal perspective, AI must include "informed consent, certification and approval as medical devices." Trustworthy AI systems must protect patient autonomy in treatment choices, and patients must consent to AI being involved in their care. Additionally, AI must not harm patients and must be continually monitored to prevent unintended harmful recommendations or misinformation. AI must also ensure access to information is equitable and culturally competent, particularly for vulnerable populations. To account for liability, AI should be required to protect data and comply with data protection regulations internationally.

## 7. Challenges and Lessons Learned

The stakes of this challenge cannot be overstated. Health misinformation powered by AI threatens to undermine decades of progress in public health, erode trust in evidence-based medicine, and create profound health inequalities. However, the same technologies that enable these threats also offer unprecedented opportunities to enhance healthcare delivery, improve patient outcomes, and strengthen health system resilience.

As AI capabilities advance and misinformation tactics become more sophisticated, the cost of delayed intervention grows exponentially in terms of health outcomes and financial burden. A single high-impact misinformation campaign can lead to surges in preventable hospitalisations, decreased vaccine uptake, and long-term public health campaigns to rebuild trust, costing governments and health systems billions annually. In contrast, investing in early detection and mitigation frameworks is comparatively cost-effective, with potential returns far outweighing the upfront implementation and coordination expenses.

The framework we have presented offers a roadmap for immediate action, but its success depends on coordinated efforts across government, industry, healthcare systems, and civil society. Currently, evaluating its performance at scale remains challenging, as the assessment process is predominantly manual and heavily dependent on human expert involvement—a resource-intensive and expensive approach. The future of healthcare—and patients' lives worldwide—depends on our collective response to this defining challenge of our time.

## 8. Conclusion

The convergence of artificial intelligence and health misinformation represents one of the most pressing challenges facing global healthcare systems today. As generative AI enable the creation of increasingly sophisticated yet potentially harmful medical content, the imperative for trustworthy, explainable AI solutions has never been more urgent. This white paper has demonstrated that the threat extends beyond individual patient harm to systemic risks to public health infrastructure, healthcare system trust, and global health equity.

While the challenges are formidable, the technological foundations for effective solutions exist. The path forward requires AI systems that are both explainable and interpretable, with transparent decision-making processes that can earn the trust of healthcare professionals and patients alike.

The ethical and legal considerations require careful navigation to ensure that solutions respect patient autonomy, promote health equity, and maintain the fundamental principles of medical ethics. Success demands attention to "autonomy, beneficence, non-maleficence and justice" throughout the development and deployment process.

The choice before us is clear: we can allow AI-powered misinformation to fragment trust in healthcare systems and jeopardise global health, or we can forge ahead with transparent, explainable AI solutions that strengthen the bonds between patients, providers, and the evidence-based medicine that serves as the foundation of modern healthcare. The path forward demands courage, collaboration, and unwavering commitment to the principles of trustworthy AI.